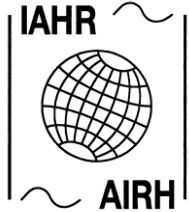

**25th IAHR International Symposium on Ice**
*Trondheim, 23 - 25 November 2020*

# A comparison of wave observations in the Arctic marginal ice zone with spectral models

**Trygve K. Løken (a), Jean Rabault (b,a), Erin E. Thomas (b), Malte Müller (b), Kai H. Christensen (b), Graig Sutherland (c) and Atle Jensen (a)**

a. Department of Mathematics, University of Oslo, Moltke Moes vei 35, 0851 Oslo, Norway
b. Norwegian Meteorological Institute, 0371 Oslo, Norway
c. Numerical Environmental Prediction Research, Environment and Climate Change Canada, Dorval, Québec, Canada
*trygvekl@math.uio.no, jeanra@math.uio.no, erinet@met.no, maltem@met.no, kaihc@met.no, graigory.sutherland@canada.ca, atlej@math.uio.no*

Increased economic activity and research interest in the Arctic raise the need for better wave forecasts in the Marginal Ice Zone (MIZ). Mathematical and numerical models of wave propagation in sea ice would benefit from more in situ data for validation. This study presents shipborne wave measurements from the MIZ where altimeter readings are corrected for ship motion to obtain estimated single point ocean surface elevation. From the combined measurements, we obtain significant wave height and zero up-crossing period, as well as one-dimensional wave spectra. In addition, we provide spectra and integrated parameters obtained from Inertial Motion Units (IMUs) placed on ice floes inside the MIZ. The results are compared with integrated parameters from the WAM-4 spectral wave model over a period of three days in the open ocean. We also compare our measurements outside and inside the MIZ with hindcast data from the new pan-Arctic WAM-3 model and the Wave Watch III (WW3) model for the European Arctic, which both model wave attenuation in sea ice. A good agreement is found with WAM-4 and WW3 in zero up-crossing period and significant wave height outside the MIZ, where deviations are less than 23%. WAM-3 is on the other hand up to 60% higher than observations. WW3 and WAM-3 are able to estimate the trends for significant wave height and zero up-crossing period inside the MIZ, although the discrepancies with respect to the observations are larger than in the open ocean. Wave damping by sea ice is investigated by looking at the spatial attenuation coefficients. Predicted attenuation coefficients are found to be 72-83% smaller for WW3 and 3-64% larger for WAM-3 compared to observations. Hence, further model tuning is necessary to better estimate wave parameters in the ice.

# 1. Introduction

The recent decline in Arctic ice cover has allowed for increased human activities in the region, which raises the importance of better forecast models and improved physical understanding of the environment to ensure safe operations (Fritzner et al., 2019). This also applies in the interface between solid ice, such as land fast ice or pack ice, and the open ocean, called the Marginal Ice Zone (MIZ). In situ wave measurements can increase our understanding of global climate systems and provide data for calibration and validation of numerical and mathematical models to describe wave attenuation in ice. However, experimental data are relatively sparse due to the harsh and dangerous environment for both researchers and instruments, combined with the inaccessibility of the regions where sea ice is present (Squire, 2007).

In this study, we present results from shipborne wave measurements in the MIZ. The methodology, first described in Christensen et al. (2013), combines a bow mounted altimeter and a motion correction instrument. We provide estimated power spectra from ocean surface elevation and integrated parameters from spectra, which are important quantities when considering wave-ice interactions. The results are compared with the spectral wave models WAM-4, Wave Watch III (WW3) and WAM-3 in the open ocean and in the MIZ, as a validation for the capability of modelling wave attenuation by sea ice. We also compare measurements in the MIZ with data from wave measuring instruments consisting of Inertial Motion Units (IMUs) placed on ice floes. From the significant wave height, the spatial damping coefficient from the observations and models are found and compared with each other.

In this paper, the data acquisition and processing methods are described in Section 2. The results are presented in Section 3 followed by a discussion in Section 4. Finally, the concluding remarks are given in Section 5.

# 2. Data and Methods

The data were obtained during a research campaign in the Barents Sea with R/V Kronprins Haakon in September 2018. Shipborne wave measurements were made continuously during cruising in the open ocean and in the MIZ, into which the ship ventured on September 19.

**Table 1.** Time, WTD and location where measurements were carried out inside the MIZ.

| Stop | Time | Position (N/E) | WTD [km] Bow | WW3 | WAM-3 |
|---|---|---|---|---|---|
| 1.1 | 04:22 | 82.126/20.736 | 0 | 5 | 0 |
| 1.2 | 06:28 | 82.246/20.245 | 28 | 34 | 15 |
| 1.3 | 08:59 | 82.355/19.803 | 61 | 58 | 31 |
| 1.4 | 12:20 | 82.436/19.674 | 82 | 65 | 53 |
| 2.1 | 13:11 | 82.421/19.579 | 77 | 65 | 51 |
| 2.2 | 14:32 | 82.359/19.544 | 65 | 61 | 32 |
| 2.3 | 15:40 | 82.294/19.389 | 46 | 43 | 27 |
| 2.4 | 16:54 | 82.228/19.275 | 18 | 27 | 20 |
| 2.5 | 18:00 | 82.163/19.183 | 7 | 12 | 7 |
| 2.6 | 19:08 | 82.099/19.046 | 0 | 0 | 1 |
| 2.7 | 20:09 | 81.994/18.982 | 0 | 0 | 0 |

Four stops were made on the way into the MIZ to deploy in situ Waves-In-Ice (WII) instruments on ice floes (Rabault et al., 2020). The sea ice concentration was visually estimated to be 10, 30, 90 and 100% for stops 1.1-1.4 respectively, and the ice thickness of the floes was approximately 1 m. Upon the return to the open ocean, a total of seven stops were made to carry out measurements unaffected by cruising speed. The location, starting time and wave travel distance (WTD) through the ice for each measurement in the MIZ are summarized in Table 1. WTD indicated "Bow" is found from the approximated ice edge from Fig. 1 and the estimated wave direction from WAM-4, while "WW3" and "WAM-3" are found by combining the ice edge and the wave direction estimated by the respective model. Prefix 1 denotes the four stops into the MIZ while prefix 2 denotes the seven stops out of the MIZ.

Figure 1(left) shows the ice edge at approximately 82.2 °N below the thin cloud cover and ship trajectory into (red) and out of (green) the MIZ. The cyan line indicates the longitudes over which the wave directions are averaged to find the mean. The ice edge is roughly recreated from the satellite image and shown in Fig. 1(right). This figure also indicates the ice edge and the mean wave direction at the ice edge at noon on September 19 estimated by the spectral models, which is further described in Section 2.2. We use the "going-to" convention for wave direction and define directions as clockwise rotation from the geographic north.

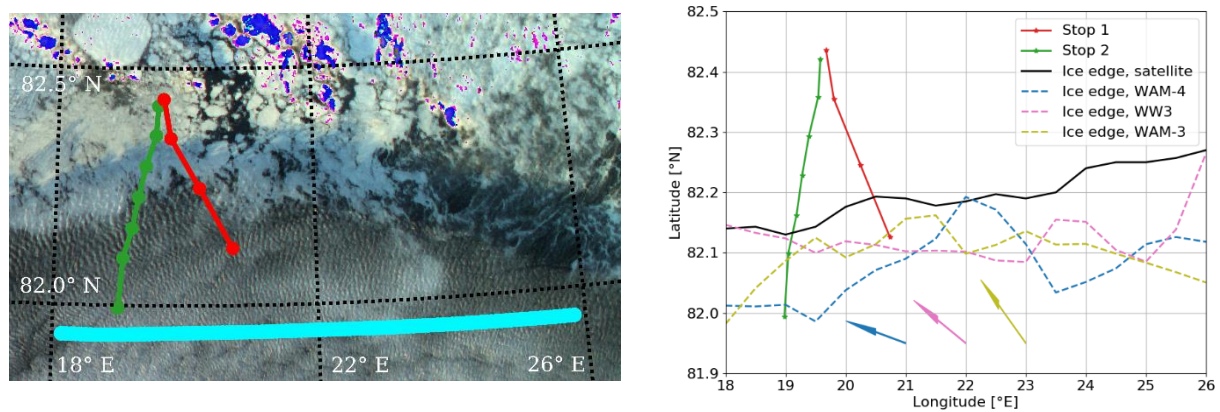

**Figure 1.** Ice edge and location of measurements. Left: Satellite image with ship trajectory into (red) and out of (green) the MIZ. The cyan line marks the averaging range for wave parameters from the models. Right: Ice edge and wave direction from models and satellite image.

The instrument setup consisted of an ultrasonic gauge (UG) mounted on a rigid pole that measured ocean surface elevation relative to the ship bow. We used a UG (Banner QT50ULB) with approximately 0.2-8 m range. The instrument emits 75 kHz ultrasonic pulses at a 10.4 Hz sampling rate. Estimated absolute surface elevation was obtained after correcting for ship motion by means of an IMU placed on deck, also in the bow section of the ship. As motion correction device, we used an IMU (VectorNav VN100). It features 3-axis accelerometers and 3-axis gyroscopes measuring at a rate of 800 Hz. After an internal Kalman filtering, the instrument gives an output frequency of 80 Hz. The gyros yield rotation angles about all three axis directly. Vertical acceleration is integrated twice to obtain ship vertical displacement about the mean. Details on the integral scheme, data filtering and other technical information on the instrument can be found in Rabault et al. (2020).

In order to obtain time series of the surface elevation, UG and IMU data at the same time instance were needed. We solved this by defining a common sampling rate of 10 Hz, which should be sufficient for resolving all relevant ocean surface features. All data were then interpolated on the common time base for the analysis. See Løken et al. (2019, Eq. 3) for details on the motion correction and estimates for the ocean surface elevation. The UG range was exceeded (saturated) at times with large wave amplitudes. We accept up to 10% saturation in a time series and will therefore discard the sample recorded at stop 2.7 in further analysis, which had a total saturation of approximately 20%.

## 2.1. Spectrum, statistical parameters and wave attenuation

Power spectra *S(f)* of the surface elevation is obtained with the Welch method, following Earle (1996). Samples are subdivided in *q* consecutive segments and ensemble averaged. Segment size is set to 200 s with 50% overlap. 20 min sampling time at 10 Hz yields a segment size of 2000 and a total number of 12000 sampling points for each measurement. A Hanning window is applied to each segment to reduce spectral leakage. Peak frequency $f_P$ is defined as the frequency at the spectral peak. The mean zero up-crossing period and the significant wave height are obtained from the spectral moments:

$$m_j = \int_{f_{min}}^{f_{max}} f^j S(f) df, \qquad [1]$$

where the cutoff frequencies $f_{min}$ and $f_{max}$ are set to 0.04 Hz and 1.0 Hz, respectively, which should include the most energetic ocean waves. Approximately the same cutoff frequencies are also applied in the spectral models. Zero up-crossing periods $T_{m02}$ are estimated from:

$$T_{m02} = \sqrt{\frac{m_0}{m_2}}. \qquad [2]$$

We investigate significant wave heights $H_S$ estimated from spectra, defined as:

$$H_S = 4\sqrt{m_0}. \qquad [3]$$

Confidence intervals for both spectra and significant wave heights are calculated from the Chi-squared distribution, following Young (1995). For spectra, the total degree of freedom (TDF) is calculated as $TDF = 2q$, and for significant wave height, TDF is found as described in ITTC (2017, p. 5). We use the Mean Absolute Percentage Error (MAPE) to compare model data with bow measurements. Systematic bias relative to observations is described with the Mean Percentage Error (MPE).

Previous field measurements indicate that waves decay exponentially in ice (Squire and Moore, 1980; Wadhams et al., 1988; Marchenko, 2018). We have investigated the attenuation of the significant wave height by fitting decreasing exponentials on the shape $H_S = Ce^{-\alpha x}$, where *C* and *α* are estimated parameters and *x* is wave traveling distance through the MIZ, by means of non-linear least squares. From the fitted curves, the spatial damping coefficients *α*, which describe wave attenuation per meter, are determined from:

$$\frac{\partial H_S}{\partial x} = -\alpha H_S. \qquad [4]$$

## 2.2. Wave models

We have investigated the performance of the three wave prediction models; WAM-4, WW3 and WAM-3 in comparison to our observations. WAM-4 does not give wave parameter where there is assumed to be an ice cover present, whereas the two latter models contain wave attenuation through the ice cover. WW3 estimates power spectra, while the WAM models only provide integrated parameters.

The WAM-4 spectral wave forecast model is run operationally by the Norwegian Meteorological Institute (Carrasco and Gusdal, 2014). Spatial and temporal resolution of the model is 4 km and 1 hour, respectively, and it runs twice a day. A hard ice boundary based on satellite images (ice concentration larger than 30%) is defined in the model.

The WAM-3 model is a pan-Arctic wave forecasting system with an effective resolution of 3 km, operated in the framework of the Copernicus Marine Environmental Monitoring System's Arctic Marine Forecasting Center. The WAM-3 wave hindcast used in the present study is forced by hourly 10 meter winds from ERA-Interim reanalysis (Dee et al., 2011) with a horizontal resolution of around 80 km. The wave spectra from the ERA-Interim are used at the lateral boundaries. The daily sea-ice concentration, ice thickness, and surface current information is taken from the TOPAZ ocean model system (Sakov et al., 2014). In areas with a sea-ice concentration larger than 20%, for the wave propagation, the frictional dissipation by the overlying ice sheet is considered as a function of the sea-ice thickness and wavenumber (Sutherland et al., 2019). Epsilon, which describes the thickness ratio of the two layers, is 0.13.

The WW3 model is a two-way coupled atmosphere-wave numerical weather prediction system in the Arctic. The atmosphere model AROME-Arctic (Müller et al. 2017) is coupled to the 3rd generation spectral wave model WW3 v5.16 using the OASIS3 model coupling toolkit. To obtain the sea-ice variables required by the wave model, we utilize the simple sea ice scheme (SICE) within SURFEX (Batrak et al., 2018). The WW3 setup used in this study uses the ST3 physics setting (equivalent to WAM4 physics) and estimates wave damping by sea ice through dissipation caused by bottom friction below a continuous thin elastic plate of ice (IC2 setting). We assume no scattering by sea ice in this framework. The coupled model is configured over the AROME-Arctic domain with a horizontal grid resolution of 2.5 km. AROME-Arctic uses 1 min. time steps while WW3 uses 5 min. time steps. The coupling frequency is every 30 min.

## 3. Results

Bow measurements are compared with model data interpolated to the location of the ship in Fig. 2(left). Valid measurements inside the MIZ, which we define as time periods where total UG saturation is below 10% and Ship Speed over Ground (SOG) is below 0.5 m/s, are highlighted with gray background color. The whole comparison spans over three days. Error statistics are summarized in Table 2, where MAPE describes the mean absolute error and MPE describes the mean error. In general, there is a good agreement between observations and model. The WAM-4 model performs best outside the MIZ compared to the bow measurements with deviations less than 23% and 8% for $H_S$ and $T_{m02}$ respectively. Deviations are approximately equally high when comparing to WW3, while WAM-3 overestimates both wave height (approx. 60%) and period (approx. 17%), especially the day before the ship ventured into the MIZ. Inside the MIZ, both models capture the larger trends when comparing them to the observations. $H_S$ is actually better predicted when the measurements are considered not valid for both models. $T_{m02}$ is more sensitive to the motion of the ship and the models are closer to the observations within the valid times.

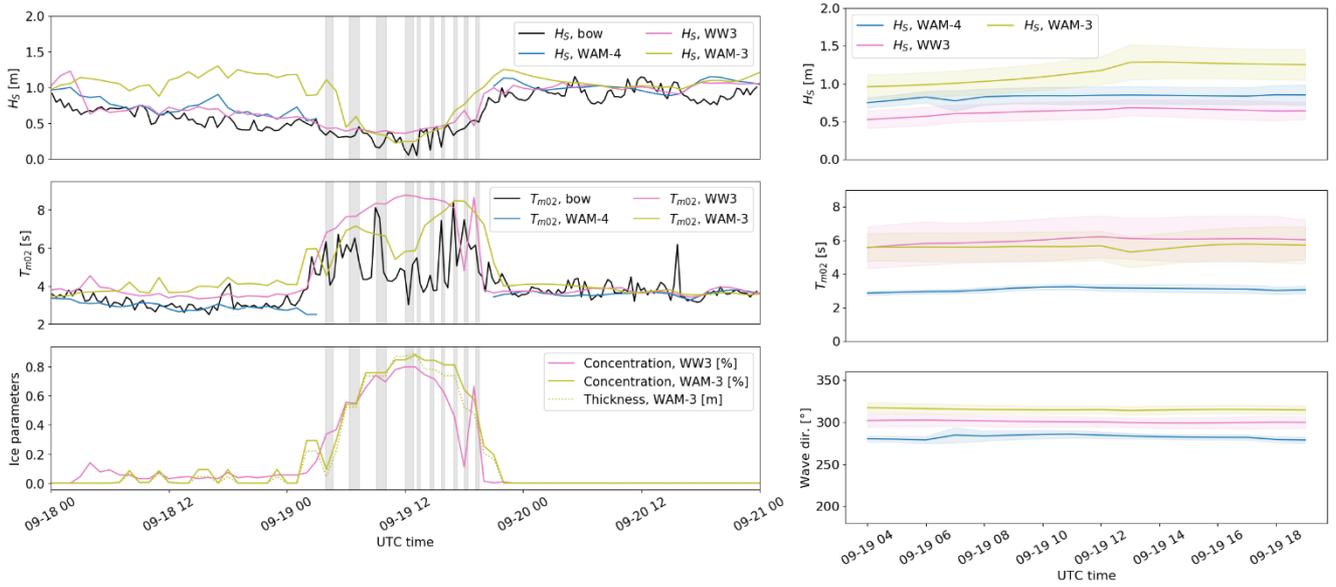

**Figure 2.** Left: long time comparison between bow measurements and the models for $H_S$ (upper), $T_{m02}$ (middle) and ice properties (lower). Measurements that are considered valid are highlighted with gray background color. Right: time series of $H_S$ (upper), $T_{m02}$ (middle) and wave direction (lower) from the wave models with spatial standard deviations shown as shaded regions. Values are extracted from locations indicated in Fig. 1.

A stationary sea state over the measurement period inside the MIZ is an advantage in our analysis. The sea state is investigated in Fig. 2(right) where time series of $H_S$, $T_{m02}$ and wave direction from the models are presented. The model data are extracted from the ice edge defined in the respective models over the range of longitudes indicated in Fig. 1(left), and the standard deviation over this range is shown as shaded areas. None of the parameters change dramatically over the time period, and the data is overall quite consistent in space.

**Table 2.** Error statistics for $H_S$ and periods $T_{m02}$ when comparing model data with bow measurements.

|  | Error | WAM-4 | | WW3 | | WAM-3 | |
| --- | --- | --- | --- | --- | --- | --- | --- |
|  |  | $H_S$ | $T_{m02}$ | $H_S$ | $T_{m02}$ | $H_S$ | $T_{m02}$ |
| MAPE [%] | Outside MIZ | 22.9 | 7.8 | 19.5 | 11.4 | 59.7 | 17.1 |
|  | MIZ, valid |  |  | 107.5 | 58.2 | 85.7 | 32.7 |
|  | MIZ, not valid |  |  | 32.8 | 53.7 | 67.2 | 39.0 |
| MPE [%] | Outside MIZ | -20.6 | 5.4 | -17.1 | -7.4 | -58.9 | -14.4 |
|  | MIZ, valid |  |  | -107.4 | -56.0 | -83.1 | 29.9 |
|  | MIZ, not valid |  |  | -32.2 | -49.6 | -61.5 | -37.3 |

Power spectra from bow measurements (black) and WII instruments (blue) with their respective 95% confidence intervals are compared to spectra from WW3 (magenta) and peak frequencies from WAM-3 (green) in Fig. 3. Stops into the MIZ are presented to the left, going successively from stop number 1.1 (upper panel) to 1.4 (lower panel) and stops out of the MIZ are presented to the right, going successively from stop number 2.1 (upper panel) to 2.6 (lower panel).

The observed power spectra from bow measurements and WII instruments in Fig. 3(left) are consistent for most frequencies for all stops in the left figure, except for stop 1.3, where the spectra deviates substantially for frequencies higher than 0.1 Hz. See Løken et al. (2019) for a discussion on this discrepancy. The spectrum from WW3 underestimates the wave energy content at stop 1.1, is consistent up to approximately 0.125 Hz at stop 1.2 and overestimates the energy at stops 1.3-1.4. The peak frequencies from WAM-3 decrease further into the ice in accordance with the observations, except for stop 1.4 where it is overestimated.

In Fig. 3(right), WW3 overestimates the wave energy content at stops 2.1-2.3 furthest into the MIZ. The model predicts a dual peak spectrum with the low frequency peak in the same frequency range as the observed spectra, but no high frequent peak was detected in the bow measurements. For stops 2.4-2.6, the spectra are comparable in strength and frequency. WAM-3 predicts an increasing peak frequency, but the growth in $f_P$ is smaller than what was observed with the bow measurements.

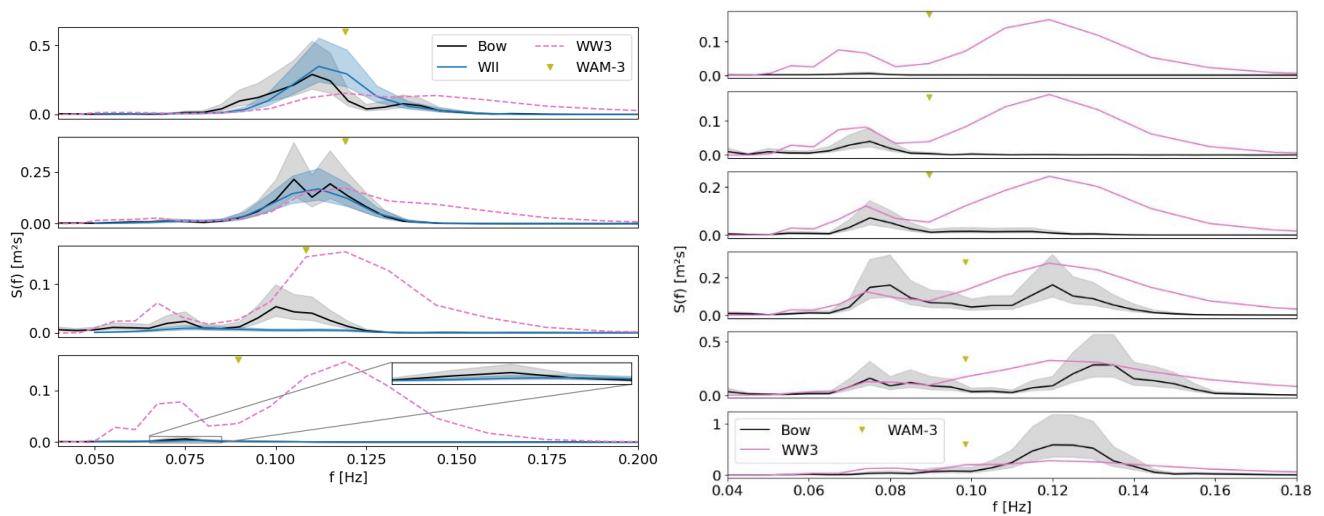

**Figure 3.** Power spectra from bow measurements (black) and WII instruments (blue) with their respective 95% confidence intervals shaded. Spectra from WW3 (magenta) and peak frequencies from WAM-3 (green). Left: going into the ice zone from upper (stop 1.1) to lower panel (stop 1.4). Right: going out of the MIZ from upper (stop 2.1) to lower panel (stop 2.6).

It is evident that waves are attenuated as they travel through the MIZ. Figure 4 shows the decrease in $H_S$ from observations (with 95% confidence intervals) and the models as function of WTD through the MIZ on the way into (left) and out of (right) the ice. Decreasing exponentials are fitted to the data and the spatial damping coefficients are found from the fitted curves with Eq. 4. The damping coefficients from the different models and observations are summarized in Table 3. The bow measurements and the WII instruments are quite consistent. WAM-3 predicts a larger damping than observations into the MIZ (attenuation coefficient 64% higher than bow measurements), but is very close to observations out of the MIZ (attenuation coefficient 3% higher than bow measurements). WW3 substantially underestimates the damping compared to observations both into and out of the MIZ (attenuation coefficient 72-83% lower than bow measurements).

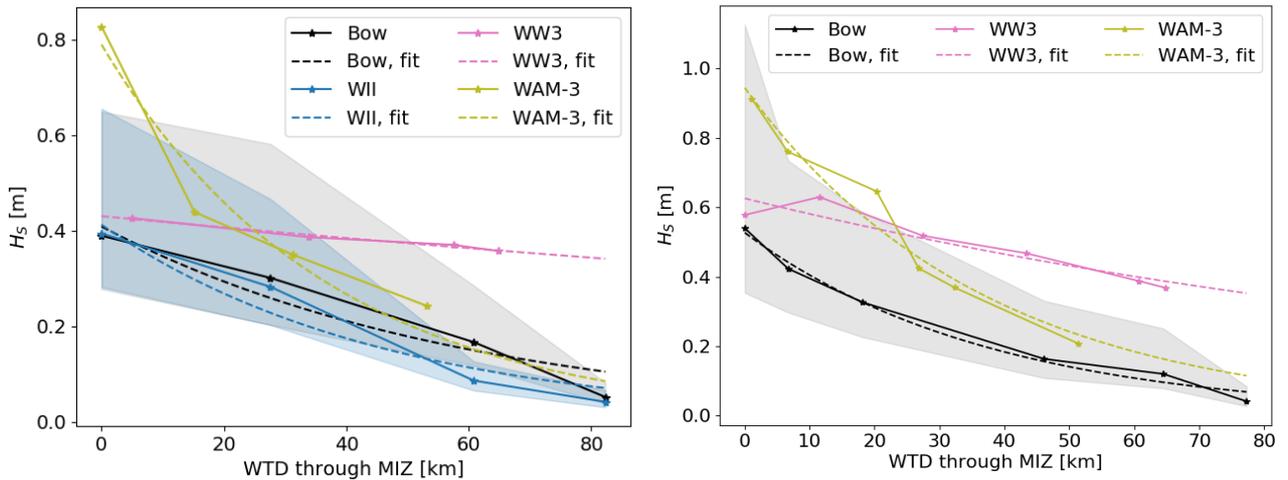

**Figure 4.** $H_S$ as function of WTD with fitted exponential decays from bow measurements (black), WII instruments (blue) with their respective 95% confidence intervals shaded, WW3 (magenta) and WAM-3 (green). Left: into the MIZ. Right: out of the MIZ.

**Table 3**. Spatial damping coefficients from observations and models into (stop 1) and out of (stop 2) the MIZ.

| $\alpha$ $[10^{-5}\ m^{-1}]$ | Bow | WII | WW3 | WAM-3 |
|---|---|---|---|---|
| Stop 1 | 1.65 | 2.15 | 0.28 | 2.71 |
| Stop 2 | 2.64 |  | 0.74 | 2.72 |

## 4. Discussion

The consistency in spectra between the bow and the WII instruments substantiates the validity of the wave measurements in the ice. The WTD through the MIZ is found from the mean wave direction and the location of the ice edge predicted by each model and therefore differs, as seen in Table 1. WAM-3 performs reasonably well compared to the observations in the MIZ when considering integrated parameters. Wave attenuation through the ice is satisfactory modeled in WAM-3 with the two-layer model of Sutherland et al. (2019), although a bit overestimated in terms of the spatial damping coefficient compared to observations. This deviation can be partly explained by the fact that the WTD found from WAM-3 is considerably smaller than the WTD found from WW3, which in again is smaller than the WTD found from WAM-4. WW3 also gives fair estimates of the integrated parameters, especially outside the MIZ, but the spectra and the damping coefficient reveals that the thin elastic plate modelling of the sea ice so far fails to estimate the attenuation of the waves.

Data recorded while a ship is in motion will contain a Doppler shifted frequency according to the wave heading angle (Collins III et al., 2017), which could be problematic in the calculation of periods. In the present study, we observe that measured $T_{m02}$ is closer to the predicted value from WAM-3, which performs best of the models inside the MIZ, when the ship is stationary. Observations of $H_S$ on the other hand, seem to be less affected by the moving ship when compared to the models, the deviations are in fact larger between models and observations at the times considered not valid. This is in agreement with the results of Cristensen et al. (2013), which reported that peak and mean periods were strongly biased by the Doppler shift in the time series, while the estimates of significant wave heights were reasonable since they only depend on the sea surface variance

## 5. Conclusions

We have presented shipborne wave measurements from the MIZ with a system combining an altimeter (UG) and a motion correction device (IMU). Our setup provides single point time series of ocean surface elevation outside of and inside the MIZ, which enables us to produce 1D power spectra and integrated parameters.

Observations have been compared with estimates from the spectral models WAM-4 (integrated parameters) in the open ocean, WW3 (integrated parameters and spectra) and WAM-3 (integrated parameters) in the open ocean and in the MIZ. We have found good agreement in zero up-crossing periods and significant wave height outside the MIZ. Predictions from WAM-4 and WW3 deviate with less than 23% over this timespan while $H_S$ predicted by WAM-3 is up to 60% off. From the time series, it is clear that the two models which predict wave attenuation through ice are able to estimate the trends for $H_S$ and $T_{m02}$ inside the MIZ, although the errors with respect to the observations are larger here. Observed $H_S$ actually matches the model prediction better when the measurement conditions are considered not valid, i.e. either when the altimeter measurement range was exceeded or when the ship was cruising. Measured $T_{m02}$ matches model predictions better during the times considered valid, most likely due to the Doppler shift induced in the time series during cruising.

Both observations and models show an exponential decay in $H_S$ through the MIZ. From the spectra and the spatial attenuation coefficients, we can conclude that WW3 underestimates the wave damping through the ice. The attenuation modeling of WW3 predicts coefficients 72-83% smaller than observations, and further tuning is therefore necessary to better estimate wave parameters in the ice. Spectra were not available from WAM-3, but the attenuation coefficients from this model match observations better and are 3-64% larger than bow measurements.

**Acknowledgments**

The authors are grateful to Øyvind Breivik for inviting us to the cruise and to the crew of RV Kronprins Haakon for their assistance. The Nansen Legacy project helped funding the campaign. Funding for the experiment was provided by the Research Council of Norway under the PETROMAKS2 scheme (project DOFI, Grant number 28062). The data are available from the corresponding author upon request.